# Introductory Physics Students in Algebra-based Courses Who Typically Worked Alone or in Groups: Insights from Gender-Based Analysis before and during COVID-19

Apekshya Ghimire * and Chandralekha Singh

Department of Physics & Astronomy, University of Pittsburgh; Pittsburgh 15260 USA.
* Correspondence: apg61@pitt.edu

**Abstract:** Collaboration with peers both inside and outside the classroom can be an invaluable tool for helping students learn physics. We investigated the impact of peer collaboration on learning physics by examining the characteristics of women and men who typically worked alone versus those who typically collaborated with peers in their algebra-based introductory physics course when they took the course before and during the COVID-19 pandemic when the classes were on Zoom. Our findings indicate that, on average, students who worked with peers had higher grades and reported greater peer influence on their physics self-efficacy during the pandemic compared to those who worked alone. We also observed that, for both women and men, a larger percentage of students typically worked in groups before the pandemic, while a greater percentage typically worked alone during the pandemic. We discuss these results in relation to students' prior academic preparation, physics grades, self-efficacy and their perception of the effectiveness of peer collaboration on their physics self-efficacy.

## 1. Introduction and Framework

Collaborating with peers to solve physics problems can often yield significant benefits for students [1-6]. Peer collaboration forms a foundation of student learning, providing opportunities for students to share knowledge and tackle challenging tasks together. Collaborative problem solving can enhance understanding of physics and boost students' confidence. However, the COVID-19 pandemic disrupted traditional in-person learning, challenging these collaborative opportunities. As classes shifted online, both students and instructors had to adapt to virtual environments to find a way to interact with each other.

**Distributed Cognition:** The distributed cognition framework explains the benefits of collaboration and support from peers based upon an increase in cognitive resources since the working memory of each individual is limited [7, 8]. Hutchins emphasized the value of distributed cognition in accomplishing intellectual tasks by enhancement of total cognitive resources.

This concept is based on the idea of sharing cognitive resources in different ways to create a cognitive system that leads to improved outcomes. He proposed that one method of achieving distributed cognition is by allowing students to collaborate in groups, thereby expanding their limited individual capacities through mutual exchange and organization of ideas and logic. When students work individually, their working memory is limited to only 5–9 chunks of information. However, when they collaborate, their combined storage capacity is significantly increased, enabling them to retain more information and, consequently, achieve better performance [1-6].

**Zone of Proximal Facilitation:** Another synergistic framework which emphasizes peer collaboration, e.g., among pairs of students, is the zone of proximal facilitation (ZPF) model [9]. This model predicts successful collaborative learning when students have partial knowledge in the subject and cannot accomplish the task alone [9]. The framework suggests that students can tackle more complex problems, which they cannot solve on their own, by collaborating with others and taking advantage of collective expertise [9]. The ZPF framework extends Vygotsky's Zone of Proximal Development (ZPD). The ZPD is defined by what students can achieve independently versus with the assistance of instructors who tailor instruction to build on students' prior knowledge [10]. The zone of proximal facilitation supports the notion that when two students working on a problem cannot solve it independently but can solve it together, the problem is out of the ZPD of each student but within the ZPD of the group. Both the distributed cognition and ZPF models emphasize the importance of peer interactions in enhancing cognitive resources and fostering a deeper understanding of physics concepts. Research indicates that students retain knowledge more effectively through group work compared to working alone [9]. These findings align with the synergistic distributed cognition framework and the zone of proximal facilitation model [11]. Collaborative work allows all collaborating students to access relevant prior knowledge and address doubts collectively via discourse. When students face difficulties solving problems individually, they can leverage their combined knowledge to find solutions. Thus, encouraging out-of-classroom collaboration can be valuable in developing a good understanding of science concepts, especially when classroom time is limited. This approach, in which students are encouraged to collaborate not only in class but also outside of class, can enable instructors to concentrate on challenging concepts beyond the zone of proximal facilitation during class, offering necessary scaffolding in those areas so that students benefit from both in-class and out-of-class activities [12].

**Positive interdependence and Individual Accountability in Peer Collaboration:** When collaborative work ensures positive interdependence [13] and individual accountability [14, 15], students at all levels of prior knowledge can benefit. This includes high-achieving students in a group, as discussion with peers and articulating one's thought processes to make them understandable to others are powerful ways to reinforce one's own understanding and clarify ideas [2, 5]. Positive interdependence is defined as a mutual reliance among all the group members [13]. It is a belief that the success of a group is only possible when everyone succeeds individually because of the group work. This is one of the most important aspects of effective collaborative learning. It creates a sense of cooperation and communication among the group members and helps them work towards a common goal [13]. Individual accountability is defined as holding each person responsible for learning in a group. Individual accountability can ensure that everyone benefits from a group task [14]. Incentive for this type of accountability can help all group members actively contribute, preventing any individual from free-riding and relying on others to complete the work. By doing so, it can maximize the engagement and effort of each member, leading to a more efficient and effective use of the group's collective cognitive resources. This not only has the potential to enhance the overall productivity of the group but can also foster a sense of shared responsibility and accountability, ultimately leading the group towards achieving their goals with greater success. Prior research shows that collaboration with peers can play a major role in enhancing students' academic performance, boosting self-efficacy, and increasing overall engagement [16-20].

**Peer Collaboration Dynamics:** The dynamics of peer interaction in academic settings significantly influence students' self-efficacy and perceptions of collaboration, particularly in the context of physics education, where peer influence on self-efficacy (PISE) plays a crucial role [21]. Some students may experience social anxiety, while others may prefer to work alone or may be compelled to do so due to various constraints [16]. Some students may not recognize the benefits of collaborative work and prioritize it. It is essential to recognize that these issues can limit their collaboration. If classroom activities do not highlight the value of teamwork or model effective peer interactions [22], students might not appreciate the importance of collaboration outside of class or know how to work well with others. In-class group work that is engaging and productive can inspire students to continue collaborating outside of class. This is particularly important for fostering a supportive environment where peer interactions can positively influence self-efficacy.

**Gender Dynamics:** Gender dynamics can also influence students' preferences for collaborative work. In STEM disciplines such as physics, there are many stereotypes about who belongs in them and who can excel in them [23, 24]. These issues can make at least some women have a lower sense-of-belonging and confidence while working with peers in groups [25, 26], leading to a preference for working alone (or working in same-gender groups even if they otherwise would like to collaborate with anyone). Furthermore, men might dominate group interactions, potentially discouraging or ignoring women's participation. Previous research studies have found that women who agree with the gender stereotype (e.g., that men are likely to be better at physics than women) perform significantly worse than those who do not agree with these stereotypes [27, 28]. Therefore, it is important to create an inclusive learning environment in physics courses where women feel comfortable to work with men and can excel academically, taking advantage of collaborative learning opportunities. Modeling effective approaches to peer collaboration in class, providing explicit encouragement to both women and men for peer collaboration both inside and outside the classroom and sharing data with students about the effectiveness of working with peers both inside and outside the class can increase their willingness and prioritization to collaborate with peers [28-32].

**Remote classes during the COVID-19 pandemic:** The COVID-19 pandemic introduced significant changes to teaching and learning practices, as institutions transitioned to remote classes. Traditional in-person classes were replaced by virtual sessions, with students attending classes on platforms like Zoom. This shift required both students and instructors to adapt to new modes of communication and interaction. Instructors had to find innovative methods to facilitate student collaboration, using tools such as breakout rooms, polls, chats, and multimedia presentations. Instructors were encouraged not to give high-stakes exams due to the stressful environment of the pandemic [33]. They were also encouraged to distribute course grades through low-stakes assessments spread throughout the semester, including quizzes, homework assignments, collaborative group projects, and take-home exams. With many students lacking social interactions with friends and family, virtual interactions offered a chance to engage in social discussions and share common experiences. While some students may have thrived in this environment, others may have found it equally challenging. Therefore, it is crucial to assess the impact of new teaching and assessment methods during COVID-19 on students' academic performance, self-efficacy and their perception of the role of peer influence on self-efficacy (PISE).

**Physics Self-Efficacy and Peer Influence on Self-Efficacy (PISE):** Some psychological constructs, e.g., students' discipline-specific self-efficacy, have been used in prior studies to understand course outcomes [34-36]. Therefore, in this study, we compare two psychological constructs during in-person versus remote physics classes: physics self-efficacy and peer influence on self-efficacy (PISE). Self-efficacy is an individual's belief in their ability to succeed in a specific activity or course [37]. Students' self-efficacy can be influenced by the classroom environment and various teaching strategies, as well as by peer interactions [38, 39]. This belief in their abilities can affect their engagement, learning and achievement in science courses [40-42]. PISE is a measure of students' perceptions of how interactions with their peers impact their confidence in physics [21]. This measure is particularly relevant in large-enrollment classes, where students tend to engage more with one another than with the instructor or teaching assistant [14]. Research has shown that higher self-efficacy correlates with higher academic achievement [42, 43]. It is also important to see how the interactions of the students with their peers influence their physics self-efficacy. While collaborative learning in diverse groups has the potential to significantly enhance academic outcomes, this benefit is dependent on creating an equitable environment that actively promotes and supports participation from all students, regardless of their gender and ethnicity [25, 44]. In such an environment, each student must feel valued and included, with their unique perspectives, and encouraged for their active participation. Ensuring equity in the classroom involves implementing strategies which provide equal opportunities for all students to engage and succeed. This might include training educators to recognize the importance of collaboration with peers, creating classroom norms that emphasize mutual respect and designing activities that ensure all voices are heard. Only by ensuring that all students, regardless of gender and ethnicity, are encouraged and supported to participate fully can the full benefits of collaborative learning be realized.

We conducted a study to investigate peer influence on self-efficacy and course grades before and during COVID-19 when classes were on Zoom, at the end of an algebra-based introductory physics course (which is a two-semester course sequence) primarily for students majoring in biosciences and/or those interested in health-related professions. We previously reported results for physics 1 in a conference proceedings paper [45] and here we expand on the research and discuss it in the context of physics 2. We also investigated students' average grade and PISE as outcomes by controlling for students' gender, their self-efficacy and prior preparation. We note that the interactions with peers related to the physics course they were asked about include students' experiences both inside and outside the class. For example, students may work together on their homework after class.

Based on the concepts of the zone of proximal facilitation and distributed cognition, we hypothesize that, in a controlled study (in which students were randomly assigned to typically work collaboratively or alone), students who engage in peer collaboration will experience greater benefits from group work, compared to those who typically worked alone (TWA) during both COVID-19 and non-COVID-19 semesters. Moreover, we predict that in a controlled study, there will be no significant differences in the performance of students who typically work in groups (TWG) between COVID-19 and non-COVID-19 semesters due to several competing affordances and constraints. For example, during the pandemic, students may experience a social oasis where peer interactions may be more highly valued due to limited opportunities for physical engagement with others. This environment may also help women navigate stereotype threats, allowing them to feel less judged and more comfortable working from home on Zoom with peers. However, the Zoom collaboration also may lack the affordances of the sharing of artifacts (materials such as textbooks or notes) and personal connections between peers that traditional in-person collaborative settings provide. Furthermore, collaboration via Zoom especially during COVID-19 could lead to scheduling conflicts and/or equity issues, particularly for students who do not have the luxury of a private space at home or had to help their family in challenging times. Considering the pros and cons of both in-person and remote collaborative learning environments, we expect no notable differences in the benefits gained from peer interaction during COVID-19 and non-COVID-19 semesters if this was a controlled study with students randomly assigned to TWA or TWG. Additionally, in a controlled study, we expect no difference in the gains in academic performance, PISE and self-efficacy between women and men typically working in groups in an equitable and inclusive environment.

Our study did not control for who typically worked with peers or alone by randomly assigning them to TWG or TWA, so students decided whether to TWG or TWA without any prompting or encouragement. In this study, we investigated several key research questions to understand the importance of collaboration on academic outcomes and self-efficacy among students before and during the COVID-19 pandemic. First, we examined whether there are differences in average grades, PISE and self-efficacy between students who typically worked alone (TWA) and those who typically worked in groups (TWG) across both time periods. Next, we explored the percentages of women and men who typically chose to work alone or in groups, along with the motivations behind their choices. We also investigated whether the patterns identified in the first research question reveal any gender differences, focusing on how women and men differ in terms of academic outcomes and self-efficacy based on whether they typically worked alone or in groups. Finally, we analyzed whether gender and TWA or TWG predict PISE and grade, controlling for self-efficacy and prior preparation, both before and during the COVID-19 pandemic. To address these research questions, we used a validated survey that included questions related to PISE and self-efficacy. Additionally, the survey asked students whether they typically worked alone or in groups along with their gender identity. This methodology allowed us to examine the relationships between PISE, self-efficacy, and academic outcomes while accounting for gender differences.

Since there was no explicit effort by the course instructors in this study to encourage group work inside or outside of the classroom and students chose to either typically work in groups or alone, this study provides baseline data on what the landscape looks like regarding students typically working alone (TWA) or typically working in groups (TWG) throughout an introductory physics course at a large research university. It can serve as a resource for instructors who encourage and model effective approaches to peer interaction by incorporating evidence-based collaborative practices in their classes.

## 2. Research Questions

We investigated the following research questions (RQs) in our study:

1. Are there any differences in average grade, peer influence on self-efficacy (PISE) and self-efficacy between students who TWA and those who TWG before and during the COVID-19 pandemic?
2. What percentages of women and men typically worked alone or in groups and what motivates them to do so?
3. Do the patterns identified in RQ1 reveal any gender differences? What are the differences between students who TWA and those who TWG among women and men?
4. Do gender and TWA/TWG predict PISE and grade when controlling for self-efficacy and prior preparation before and during COVID?

## 3. Methodology

### 3.1. Participants and Courses

The study was conducted at a large public research university in the U.S. and focused on students enrolled in the algebra-based introductory Physics 2 course, which covers topics like electricity, magnetism, circuits, and physical optics. We surveyed students over three consecutive spring semesters, asking whether they typically worked with peers or alone, and asked them about psychological factors such as self-efficacy and peer influence on self-efficacy (see Table 1). Two of the semesters occurred before the COVID-19 pandemic, while one took place during the pandemic. During the pandemic semester, all classes were conducted remotely via Zoom with no in-person meetings. Consequently, students collaborated online, creating a group dynamic different from traditional in-person classes where direct interaction was possible. The survey was administered during the final week of mandatory recitations led by teaching assistants. For analysis, we included students who completed the survey and passed an attention check (requiring the selection of option "C"). Depending on the instructor, students received either extra credit or a participation grade for completing the survey. Most respondents were juniors and seniors majoring in biological sciences or health-related fields. A total of 755 students completed the survey before COVID-19, while 423 students completed the survey during the COVID-19 semester when classes were held remotely via Zoom. Although remote instruction was used due to the pandemic, the university typically offers only in-person physics courses. Demographic data such as age, gender and ethnicity or race were provided by the university through an honest broker process, ensuring the research team could access this information without knowing participants' identities.

**Table 1.** The peer influence on self-efficacy (PISE) and self-efficacy (SE) items on the survey.

| Item No. | My Experiences and Interactions with Other Students in This Class: |
|---|---|
| PISE 1 | Made me feel more relaxed about learning physics. |
| PISE 2 | Increased my confidence in my ability to do physics. |
| PISE 3 | Increased my confidence that I can succeed in physics. |
| PISE 4 | Increased my confidence in my ability to handle different physics problems. |
| SE 1 | I am able to help my classmates with physics in the laboratory or in recitation. |
| SE 2 | I understand concepts I have studied in physics. |
| SE 3 | If I study, I will do well on a physics test. |
| SE 4 | If I encounter a setback in a physics exam, I can overcome it. |

*3.2. Prior Academic Performance*

Prior academic preparation was assessed using high school grade point averages (HS GPAs) and Scholastic Assessment Test (SAT) or American College Testing (ACT) scores (which are often required or recommended by many US colleges and universities). HS GPAs ranged from 0 to 5, and students with GPAs above 5 (about 1% of the sample) were excluded from the analysis due to the likely use of a different grading system at their schools. The SAT scores tests students' skills in reading, writing and math, with each section ranging from 200 to 800 [46]. It was used along with converted American College Testing (ACT) scores, based on established conversion tables [47]. The ACT covers four sections: English, math, reading and science, with an optional writing section. Each section is scored on a scale of 1 to 36, and the composite score is an average of the four section scores [46]. We only focused on SAT/ACT math scores in this paper. Both demographic and prior academic data were obtained from de-identified university records through an honest broker, ensuring confidentiality and consistency in the analysis.

*3.3. Physics Grade*

Instructors provided academic performance metrics in the form of the final course grade, which was based upon the performance on homework, quizzes, midterm exams and final exams. The final grade represents the overall performance of a student in the physics course. We obtained final grades for each course from the university and linked them to students' gender and survey responses through an honest broker process. Course grades were assigned on a 0–4 scale, with A = 4, B = 3, C = 2, D = 1, F = 0 or W (late withdrawal). The suffixes '+' and '−' adjust the grade points by 0.25 (e.g., B−= 2.75 and B+ = 3.25), except for A+, which is reported as 4 similar to A.

*3.4. Survey*

Two psychological constructs that played an important role in determining the performance of students in physics courses were studied. The first one is peer influence on self-efficacy (PISE), which measured whether students thought that working with their peers was beneficial to their confidence to do physics [5]. The second one is physics self-efficacy (SE), which is a student's belief in their ability to complete a physics-related task [37]. The items in the study were designed on a Likert scale of 1–4 (1—Strongly Disagree, 2—Disagree, 3—Agree, 4—Strongly Agree) [48, 49]. A lower score was indicative of a negative endorsement, while a higher score was related to a positive endorsement. The items for PISE and physics self-efficacy [20, 21, 50] were adapted from previously validated surveys [26, 48, 51]. We also performed confirmatory factor analysis (CFA) [52] to confirm survey validity. The Comparative Fit Index (CFI) and Tucker–Lewis Index (TLI) were 0.95, the Root Mean Square Error of Approximation (RMSEA) was 0.08 and the Standardized Root Mean Residual (SRMR) was 0.06. Cronbach's alpha was used to measure the internal consistency of the items. Cronbach's alpha for PISE was 0.92 which is considered excellent [53]. For SE, the Cronbach's alpha was 0.78 which is considered reasonable [53].

All PISE and physics self-efficacy survey items can be found in Table 1. To ensure we were measuring domain-specific psychological constructs, we explicitly mentioned physics in the survey items, such as 'I understand concepts I have studied in physics'. An ideal course outcome is that all students have high self-efficacy and high PISE. The instructors did not offer any incentives for collaboration with other students. As a result, students chose to collaborate on their own initiative, and their responses are self-reported based on these interactions both inside and outside the classroom.

We inquired about the students' typical mode of work through the following question in the survey: Most typically in this physics course...
   A. I worked alone
   B. I worked with students mostly of my own gender.
   C. I worked in mixed-gender groups.

We grouped students who selected option B and option C into the category of students who typically worked in groups (TWG). Those who selected option A were classified as students who typically worked alone (TWA). Thus, the word "typically" refers to the students' primary mode of working, either alone or in groups. While they may have occasionally worked in different settings (alone or in groups), their predominant working style is defined by what is considered typical for them. This was followed by an additional question: "In one sentence, summarize why you primarily worked alone, in same gender groups, or mixed gender groups". We inquired about the reasons why students typically worked alone or in groups in the survey using an open-ended question. We do not have any reasonings from the students during COVID-19 as all these reasonings were collected in the semesters after the pandemic once the students were back to the traditional classroom settings. So, these reasonings may or may not be similar for the students who TWA or TWG before or during the pandemic. We used ChatGPT-3.5, a large language model, to summarize all the reasons collected from the students and identify the top five reasons for the students to typically work alone or in groups along with the gender differences. Both researchers reviewed the reasoning data to ensure the validity of these outcomes and confirmed that these were indeed the most common reasonings from the students.

*3.5. Analysis*

To prepare for analysis, PISE, SE and student grades were standardized. For each student, the average of the four PISE-related items (see Table 1) was calculated and converted to z-scores using the formula $Z = (X − \bar{X})/\sigma_X$, meaning that observations were converted to measure the number of standard deviations (SD, $\sigma_X$) from the mean of all students ($\bar{X}$) and rescaled to a range from 0 to 1 using $(Z − Z_{min})/(Z_{max} − Z_{min})$ [52]. The z-scoring was performed separately for non-COVID-19 and COVID-19 semesters. This process was repeated for grades and self-efficacy items. A higher score indicates a higher grade, higher PISE or higher SE.

To assess gender differences and variations in grades, PISE and self-efficacy between students who typically worked alone (TWA) and those who worked in groups (TWG), we conducted unpaired *t*-tests and calculated Cohen's *d* for effect size. We recognize that gender is not a binary construct, but the university data only included binary categories of women and men. Cohen's *d* was calculated using $d = (\bar{X}_1 − \bar{X}_2)/S_{pooled}$, where $\bar{X}_1$ and $\bar{X}_2$ refer to the sample means of these two groups and $S_{pooled}$, the pooled standard deviation [52], is a measure of effect size. We used the following standards: small *d* ~ 0.2; medium *d* ~ 0.5; and large *d* ~ 0.8. We also investigated the statistical significance of these results using a p-value, where $p < 0.05$ is considered as a statistically significant result.

Multiple regression analysis was used to explore the predictive relationship between PISE and grades, considering other variables such as gender, HS GPA, SAT math scores and self-efficacy. For each regression model, we reported the standardized *β* coefficients, sample size and Adjusted R-squared. Adjusted R-squared measures variance while accounting for the number of variables, making it suitable for multiple regression models with many predictors. Standardized coefficients were used for their comparability in units of standard deviation [52].

## 4. Results and Discussion

To answer RQ1, we investigated the differences in grades, peer influence on self-efficacy (PISE) and overall self-efficacy between students who typically worked alone (TWA) and those who typically worked in groups (TWG), without considering gender. Initially, our analysis revealed no statistically significant differences in grades between these groups prior to the COVID-19 pandemic (see Figure 1, with statistically significant values bolded). However, during the pandemic, students who worked in groups earned statistically significantly higher average grades. The effect size for this difference was medium, with *d* ~ 0.42 for grades. As shown in Figure 1, grades for students who worked alone or in groups increased during the pandemic.

Although we predicted that there would be no difference in the performance of the students who TWG before and during COVID-19, this improvement in grades among those who collaborated with others during the pandemic could be attributed to the introduction of low-stakes assessments distributed throughout the semester, which likely reduced the pressure of a final cumulative exam and facilitated better performance. The difference in PISE scores between the students who TWA and TWG was statistically significant both before and during the COVID-19 pandemic. Although the magnitude of this difference falls within the range of small effect sizes, it increased from 0.21 to 0.29 during the pandemic. For physics self-efficacy, the analysis showed that the difference between the two groups was minimal, with an effect size of just 0.01, indicating statistical insignificance. This suggests that there is no statistically significant difference in working alone versus in groups on students' self-efficacy, irrespective of the time period studied (COVID-19 vs. before COVID-19).

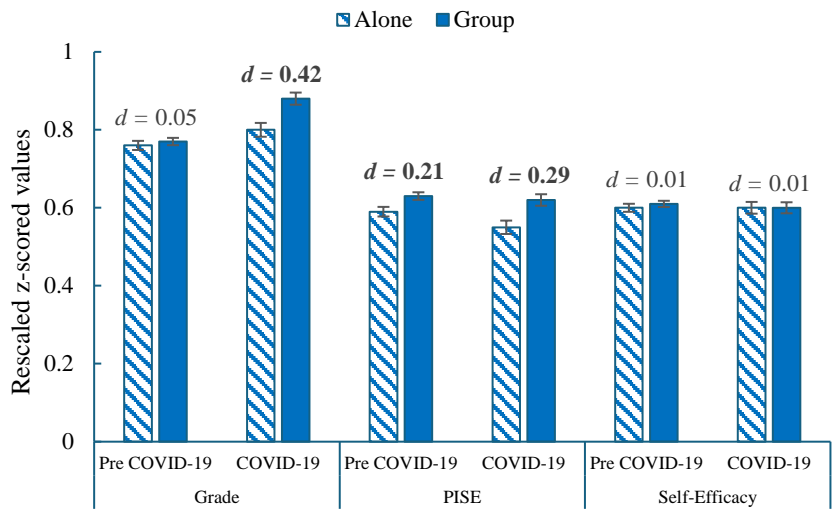

Figure 1. Grade, PISE and SE of students who TWA (alone)/TWG (group) before and during the pandemic (statistically significant values are bold).

We also investigated the differences among the students who TWA before and during COVID-19 and the students who TWG before and during COVID-19, as seen in Figure A1 in Appendix A. The only significant finding was that students who TWG earned higher grades during the pandemic, with no significant differences in PISE and physics self-efficacy. Additionally, students who TWA showed a statistically significant difference in grades before and during COVID-19, with higher grades observed during the pandemic. However, the improvement in grades was more pronounced for students who TWG compared to those who TWA. This could be because our study was not controlled and relied on self-reported data from the students. The lack of significant changes in PISE and physics self-efficacy align with our hypothesis that there would be no significant differences for students engaging in peer collaboration before and during COVID-19.

For RQ2, the top reasons students gave for why they chose to work alone or in groups were the same—to enhance their understanding of physics and improve their interpretation of physical laws. We observed that more students typically worked in groups (TWG) before COVID-19, as shown in Table 2. Many students who worked with others noted that being friends or living in the same dormitory facilitated group work. Those who preferred group work primarily cited the benefits of concept clarification through discussions and debates. One student stated, "I was struggling working entirely by myself so I reached out to other students who had taken the course". Another student said, "It is easier to work with other people to understand problems". They highlighted that group work fosters collaboration and the exchange of ideas, leading to a deeper understanding and better problem-solving. Additionally, students appreciated the social support and motivation that group work provides, which helps them stay engaged and committed to solving problems. One student stated, "The guys I sat with were legends and made a hard class very bearable".

Table 2 indicates that before the pandemic, a greater percentage of men (47%) worked alone compared to women (33%). This percentage increased during the COVID-19 period, with 63% of men and 51% of women typically working alone.

Both women and men who preferred working alone mentioned that it allowed them to concentrate better on tasks. They also noted the advantage of exploring challenging topics in depth and making mistakes without the fear of peer judgment. Efficiency and the ability to work at their own pace were also key factors influencing their decision to work alone. We observed some gender differences in the reasons given for working alone. Many women cited feelings of intimidation by others' knowledge (some specifically mentioned men in this context), which could be related to self-efficacy and/or social dynamics. One woman explained, "I mainly worked alone because working with other people causes me to overthink my intelligence and how academically capable I am. I would rather learn the content by myself then feel pressured and uncomfortable working with someone else". Some women also reported feeling discouraged by male peers, leading them to work alone. Another woman mentioned, "I felt that the boys in class were overly confident in their abilities and didn't understand why someone might not know". In contrast, men who preferred working alone often mentioned their comfort with independent problem-solving or their confidence in their own abilities. One man said, "I find it hard to rely on people other than myself", while another noted, "I would say for the most part my work was done independently in proximity with others of mixed gender. I wouldn't ask many questions from others as I was mostly confident in my physics knowledge".

**Table 2.** Number of women and men who worked alone and in groups before and during COVID-19 pandemic, along with the corresponding chi-squared value showing statistically significant results.

|  | **Before COVID-19** | | **COVID-19** | |
| --- | --- | --- | --- | --- |
|  | **Women** | **Men** | **Women** | **Men** |
| Alone | 152 (33%) | 135 (47%) | 133 (51%) | 102 (63%) |
| Group | 315 (67%) | 153 (53%) | 129 (49%) | 59 (37%) |
| $\chi^2$ | 15.52 ($p < 0.001$) | | 6.4 ($p = 0.011$) | |

Additionally, students cited scheduling conflicts as a factor, noting that working according to their own schedules was easier than coordinating with others. Some students also mentioned not knowing anyone in the class or struggling to make friends, which led them to study alone by default. Gender differences were also evident, with some men feeling that their approach to the subject might be different, e.g., too intense, for their female peers. One man stated, "I find that I am too intense for people of the other gender to keep up with me". This implies that these men felt that their intensity and pace in discussing physics might be difficult for women to keep up with. These types of comments may point to the importance of creating more inclusive and supportive learning environments where all students, regardless of gender, can succeed and collaborate effectively.

Regarding RQ3, the gender difference in grades among students who collaborated with others (TWG) was statistically significant before COVID-19 but became statistically insignificant during the pandemic (see Figure 2). The effect size of the difference between men and women who TWG was small, at $d$~0.22 before the pandemic. Figure 2 indicates that men who TWG had higher grades than women who TWG before the pandemic, suggesting that men had a higher correlation between grade and peer collaboration compared to women. We observed that women who collaborated with peers during COVID-19 obtained higher grades than those who collaborated before the pandemic. In fact, women who TWG during COVID-19 earned higher grades than men who TWG; however, the difference was not statistically significant. In future investigations, it would be valuable to ask women in the interviews whether the peer collaboration in a remote learning environment offered a consistent experience across genders and helped overcome stereotype threat. For those who TWA, the effect size for gender difference is relatively small, $d$~0.07 before COVID-19. But the gender differences for those students who TWA during the pandemic is statistically significant with a medium effect size, of $d$~0.37.

There was no statistically significant difference in grades between those who TWA and those who TWG for both men and women before the pandemic. The difference was statistically significant during COVID-19 with a medium effect size, $d$~0.60, for women, but it was not statistically significant for men with a small effect size, of $d$~0.18 (see Figure A2 in Appendix B). This shows that TWA/TWG did not have a role in predicting grades for men during COVID-19, but it played an important role for women, with higher grades among those who worked with their peers.

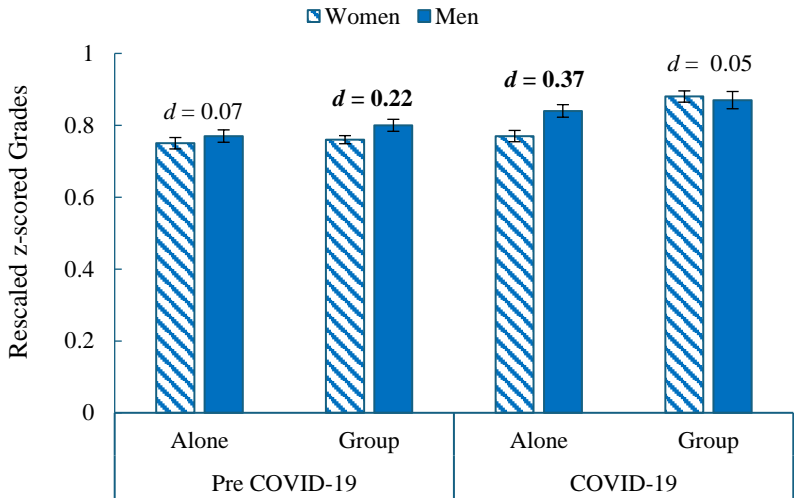

Figure 2. Grades of the students who TWA (alone)/TWG (group) divided by gender (statistically significant values are bold).

The gender difference in PISE is statistically significant for students who TWA and those who TWG before the pandemic, but the difference is smaller for the students who typically worked with their peers with a small effect size, $d\sim0.20$ (see Figure 3). During COVID-19, we see that the gender difference is not statistically significant among those who worked alone and those who collaborated with others, with higher PISE among men compared to women. If we compare women who TWA and TWG, we see that those who typically collaborated with others have a higher PISE (see Figure A3 in Appendix B). The effect size for this difference is medium, $d\sim0.30$, before COVID-19, and it is statistically significant. Women who collaborated with others have significantly higher PISE during COVID-19 as well with a medium effect size, of $d\sim0.32$. This result makes sense as individuals who perceive benefits in collaborating with others are likely to TWG and have higher levels of PISE. For men, we find that the difference is statistically significant during COVID-19 with a medium effect size, of $d\sim0.32$, but it is not statistically significant before COVID-19 with a small effect size, of $d\sim0.16$.

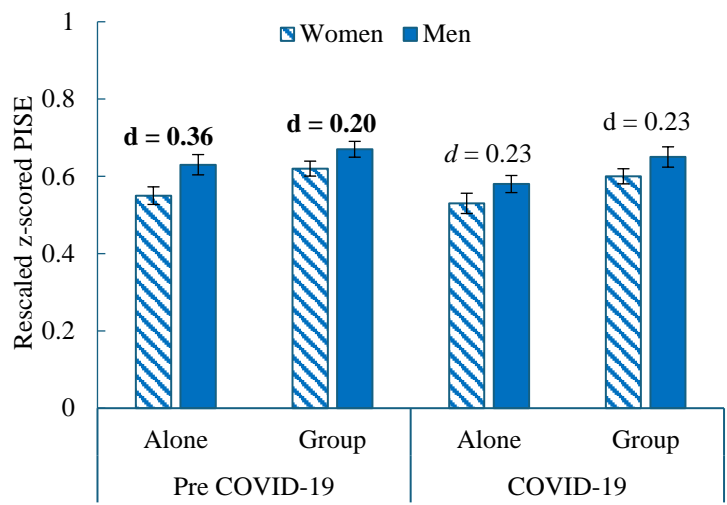

Figure 3. PISE of the students who TWA (alone)/TWG (group) divided by gender (statistically significant values are bold).

We observed that the gender difference in physics self-efficacy is higher for students who TWG than those who TWA before the pandemic (see Figure 4). It is important to note that this is not a controlled study; students chose to TWA or TWG and the physics self-efficacy values are self-reported by those students. The experiences of the students during their initial semester (or even before coming to college) could set a baseline for their self-efficacy, which may increase or decrease as they progress to this course. However, the gender difference among students who TWG not only decreases but also becomes statistically insignificant during the pandemic. The correlation between group work and the reduction in the gender gap are noteworthy. We also observed that there was no statistically significant difference between women who TWA and TWG, with a small effect size both before and during the pandemic (see Figure A4 in Appendix B). This pattern was also observed among men with a small effect size, of $d\sim 0.05$ before and $d\sim 0.17$ during the pandemic.

Regarding RQ4, Table 3 shows the results of some useful regression models predicting average physics grade. The independent variables (predictors) used for regressions are in the first column and any blank spaces indicate that the predictor in the corresponding row is not used in that model. The strength of each predictor is given by the standardized coefficients ($\beta$) controlling for all other variables. For each change in a standard deviation shown in a predictor variable, the model predicts a change in standard deviation in the outcome variable controlling for all other predictor variables. There were several regression models for grades, but after thorough discussion between both researchers, we have chosen to present only those that show significant differences and are most relevant. We included all predictors used in the analysis but only showcased the models where the predictors indicate a significant difference when other variables are included or excluded.

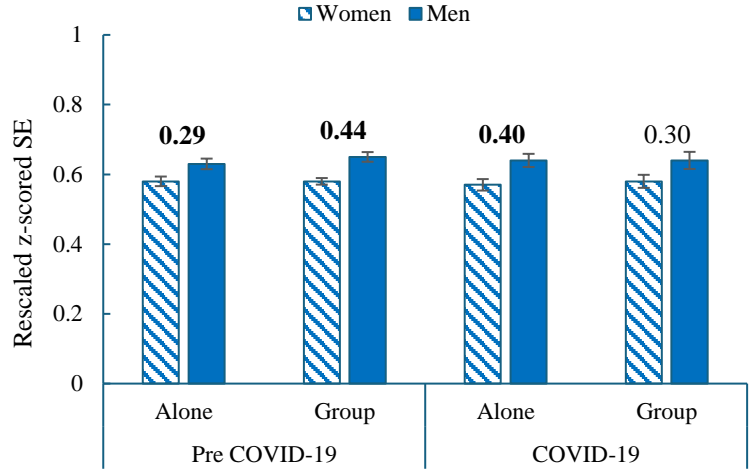

Figure 4. Self-efficacy of the students who TWA (alone)/ TWG (group) divided by gender (statistically significant values are bold).

Table 3. Standardized coefficients of regression models predicting grade. SE = physics self-efficacy. Significant predictors are bold. * Represents $p < 0.05$, ** represents $p < 0.01$, and *** represents $p < 0.001$ [52].

| Before COVID-19 (Grade as the Outcome Variable) | | | | |
|---|---|---|---|---|
| Predictors | Model 1 | Model 2 | Model 3 | Model 4 |
| Gender | **0.07 *** | 0.02 | 0.04 | 0.03 |
| Group | 0.03 | 0.02 | 0.01 | 0.02 |
| PISE | | **0.09 *** | **0.18 **** | **0.10 ** ** |
| HS GPA | | | **0.23 **** | **0.23 **** |
| SAT Math | | | **0.35 **** | **0.34 **** |
| SE | | **0.24 **** | | **0.14 **** |
| Adj. | 0.004 | 0.09 | 0.28 | 0.29 |
| COVID-19 (Grade as the Outcome Variable) | | | | |
| Predictors | Model 1 | Model 2 | Model 3 | Model 4 |
| Gender | 0.09 | 0.06 | **0.09 *** | **0.07 *** |
| Group | **0.22 **** | **0.21 **** | **0.16 ** ** | **0.17 ** ** |
| PISE | | 0.02 | **0.15 ** ** | 0.06 |
| HS GPA | | | **0.13 ** ** | **0.14 **** |
| SAT Math | | | **0.28 **** | **0.27 **** |
| SE | | **0.20 ** ** | | **0.15 ** ** |
| Adj. | 0.05 | 0.09 | 0.19 | 0.20 |

The regression models presented in Table 3 provide a detailed analysis of the predictors of student grades, both before and during the COVID-19 pandemic. The regression models in Table 3 indicate that neither gender nor typically working alone or typically working in a group strongly predicts student grades before COVID-19. However, physics self-efficacy emerges as a strong predictor in model 2, followed by PISE, though this model explains only a small amount of variance in grades. Models 3 and 4 highlight that high school GPA and SAT math scores significantly influence student course grades. During COVID-19, working alone or in a group also becomes a strong predictor of course grades, indicating that students who typically worked with others had significantly higher grades than those who worked alone. This indicates a shift in the factors influencing academic performance, likely due to the altered learning environments and the increased importance of peer interactions and support during remote or hybrid learning settings. Prior preparation and physics self-efficacy remain strong predictors of grades during COVID-19, but the variance explained by these models is lower compared to the models before COVID-19, as shown in Table 3. This suggests that the pandemic introduced additional variables that affected the student grades, which are not fully accounted for by the predictors included in these models.

Table 4 presents the regression models for PISE both before and during COVID-19. It is evident that men exhibited a higher PISE than women when controlling for whether they typically worked alone or in groups (Model 1) or when controlling for both typically working alone or in groups and their grades (Model 2). However, both models explain only a small amount of variance, as indicated by the Adjusted $R^2$ values not exceeding 0.08. This suggests that additional factors are needed to explain this difference. Model 3 in Table 4 suggests that physics self-efficacy is a strong predictor of PISE, explaining 32% of the variance before pandemic and 38% of the variance during pandemic. When physics self-efficacy is included in the analysis (Model 4), both gender and grade become non-significant predictors during the pandemic. Model 4 explains the most variance among the four regression models both before and during COVID-19. This indicates that physics self-efficacy may account for the differences in PISE related to gender and grade observed in other models. This finding suggests that creating a learning environment in which all students have high self-efficacy (particularly women) could be valuable in physics courses.

**Table 4.** Standardized coefficients of regression models predicting PISE. SE = physics self-efficacy. Significant predictors are bold. (* represents $p < 0.05$, ** represents $p < 0.01$, and *** represents $p < 0.001$) [52].

| Before COVID-19 (PISE as the Outcome Variable) | | | | |
|---|---|---|---|---|
| Predictors | Model 1 | Model 2 | Model 3 | Model 4 |
| Gender | **0.13 ** | **0.11 ** | | 0.02 |
| Alone/Group | **0.12 ** | **0.11 ** | | **0.09 ** |
| Grade | | **0.22 ** | | **0.06 *** |
| SE | | | **0.57 *** | **0.55 *** |
| Adj. | 0.02 | 0.07 | 0.32 | 0.34 |
| COVID-19 (PISE as the Outcome Variable) | | | | |
| Predictors | Model 1 | Model 2 | Model 3 | Model 4 |
| Gender | **0.11 *** | **0.09 *** | | 0.01 |
| Group | **0.16 ** | **0.13 *** | | **0.15 *** |
| Grade | | **0.15 ** | | 0.01 |
| SE | | | **0.62 *** | **0.62 *** |
| Adj. | 0.03 | 0.05 | 0.38 | 0.40 |

**5. Summary, Conclusions and Future Directions**

In this study, we analyzed baseline data from a period when no explicit efforts were made to encourage group work among students, either inside or outside the classroom, nor were students randomly assigned to TWA or TWG. The results presented here provide a valuable reference point for future research on the effectiveness of peer collaboration in educational settings.

Regarding RQ1, in the absence of incentives for participating in group work or randomized assignment into TWA or TWG, our findings revealed that students who TWG achieved significantly higher grades during the COVID-19 period compared to those who TWA. While this is not a controlled study, group work could be an effective strategy that instructors can utilize in their classes to improve students' academic outcomes, consistent with our framework of distributed cognition and zone of proximal facilitation. These students also had higher PISE both before and during the pandemic in comparison to those who TWA. Students who TWG earned better grades during the pandemic than they did prior to it, with no significant changes in PISE or self-efficacy before and during COVID-19. Thus, the better grades of those who TWG during the pandemic goes against our null hypothesis if this were a controlled study. In particular, students typically working with peers benefited more from collaboration during the pandemic than those in the in-person classes even though it may have been difficult to share opinions on Zoom. In contrast, students who TWA showed a statistically significant increase in grades from before to during the pandemic.

Regarding RQ2, we find that a higher percentage of students TWG before COVID-19, while a higher percentage of students TWA during the pandemic. Among women, the most striking differences are that one-third of the women TWA (and two-thirds TWG) before the pandemic, but during the pandemic, the percentages of women who TWA and TWG are relatively similar. Among men, the most striking differences are that the percentages of men who TWA and TWG are relatively similar before the pandemic, but two-thirds of the men TWA (and one-third TWG) during the pandemic. The primary reason for group work was that friendships and living arrangements fostered collaboration, along with the social support and motivation it offered. Conversely, the main reason cited for working alone for women was that many women felt intimidated by their peers' knowledge, especially men, which impacted their self-efficacy. In contrast, men expressed a greater comfort with independent problem-solving and felt confident in their abilities.

Regarding RQ3, our findings show that women who collaborated with peers during COVID-19 achieved higher grades than those who worked with others before the pandemic. In fact, women who TWG during COVID-19 outperformed men in the same category, although this difference was not statistically significant. However, the gender differences in grades among students who TWA were statistically significant only during the pandemic, not before. When we compared the grades of women who TWA with those who TWG, as well as men who TWA and TWG before and during COVID-19, the only significant difference was observed between women who TWA and TWG during the pandemic. For PISE, we observe that men had a significantly higher PISE compared to women whether they TWA or TWG before the pandemic. One hypothesis for men's higher PISE compared to women's is men's greater initial self-efficacy and women not being able to take full advantage of peer interaction due to gender stereotypes. However, women who TWG had significantly higher PISE than those who TWA, both before and during the pandemic. This is understandable, as individuals who value collaboration are more likely to engage with others and, as a result, have higher PISE. For men, those who TWG had significantly higher PISE scores compared to those who TWA only during the pandemic. The gender difference in self-efficacy among students who TWA and those who TWG was highly significant before the pandemic; however, it was only significant among students who TWA during the pandemic. This reduction in gender difference can be attributed to the enhanced self-efficacy observed among women who TWG during the pandemic. However, when we compared the students who TWA and TWG within each gender, we found no statistically significant differences in self-efficacy across any of the groups. The higher PISE among men and the significant gender differences in self-efficacy may highlight inequities and the necessity of establishing an equitable and inclusive learning environment in physics courses, ensuring that women can benefit from group work just as much as men do.

Regarding RQ4, our findings suggest that students' prior preparation, measured by high school GPA and SAT math scores, along with their physics self-efficacy, are the strongest predictors of their physics grades. Additionally, physics self-efficacy is a strong predictor of PISE. Students who have better prior preparation, including higher high school GPAs and SAT math scores, along with high self-efficacy, tend to achieve higher grades in physics. Therefore, it is crucial to focus on these issues to improve their academic performance.

Regarding future directions, during the pandemic, many students experienced isolation due to lockdowns and social distancing measures, which impacted their academic and social lives. To better understand student behavior and its impact on learning, conducting in-depth interviews with students can provide valuable insights into the benefits of peer collaboration in both online and traditional in-person classroom settings. These one-on-one interviews could explore how social interaction and opportunities for peer collaboration were perceived as valuable (or not) when students interacted on Zoom during the pandemic or in their in-person classes. These interviews could explore gender-specific experiences and differences in self-efficacy and academic performance, as well as identify the technological and logistical challenges encountered during both online and in-person collaboration.

An important aspect to investigate in these interviews is the extent to which Zoom classes and peer interactions during the pandemic provided an environment that reduced the stereotype threat that many women experience in traditional in-person physics classes. The physical separation and the option to turn off their video in Zoom classes may have allowed some women to focus better on collaborative work without worrying about being judged by others. If this hypothesis is true, it could explain the improved engagement and performance observed among women who participated in collaborative work during the pandemic. This would suggest that creating environments where students feel less stereotyped and more comfortable can enhance their participation and success in collaborative settings. Future research can delve deeper into this aspect to determine whether online classes are truly effective in reducing stereotype threat among women.

Understanding these gender dynamics can provide valuable insights into how different settings impact the motivation and engagement of both women and men in collaborative work. Such information can help instructors design more effective collaborative learning environments that enhance learning outcomes for all students. By exploring students' perceptions and experiences through interviews, we can gain a deeper understanding of the factors that contribute to successful peer collaboration and how to create more equitable and inclusive educational spaces. This, in turn, can inform teaching practices and support the development of strategies that maximize the benefits of peer interactions.

Instructors play an important role in encouraging collaboration by creating a supportive classroom where students feel comfortable sharing ideas and learning from each other. In appropriate contexts, grouping students with different strengths can help them see new perspectives, improving their understanding and engagement. Instructors should also encourage collaboration outside of class, such as for homework and projects to extend the benefits of teamwork. They can assign roles to make sure that everyone contributes equally to the group and conduct individual assessments to hold the students accountable. It is also important that the instructors teach important teamwork skills like communication and conflict resolution to make group work more effective. Instructors can motivate students by sharing research that shows how group work can lead to better grades, higher confidence and deeper learning. Using tools like shared documents or group chats can further support collaboration, especially in online or hybrid classes. This approach can help create a more inclusive and supportive learning environment, making sure all students, regardless of gender, benefit from working together.

**Funding:** This research was funded by National Science Foundation grant DUE-1524575.

**Appendix A**

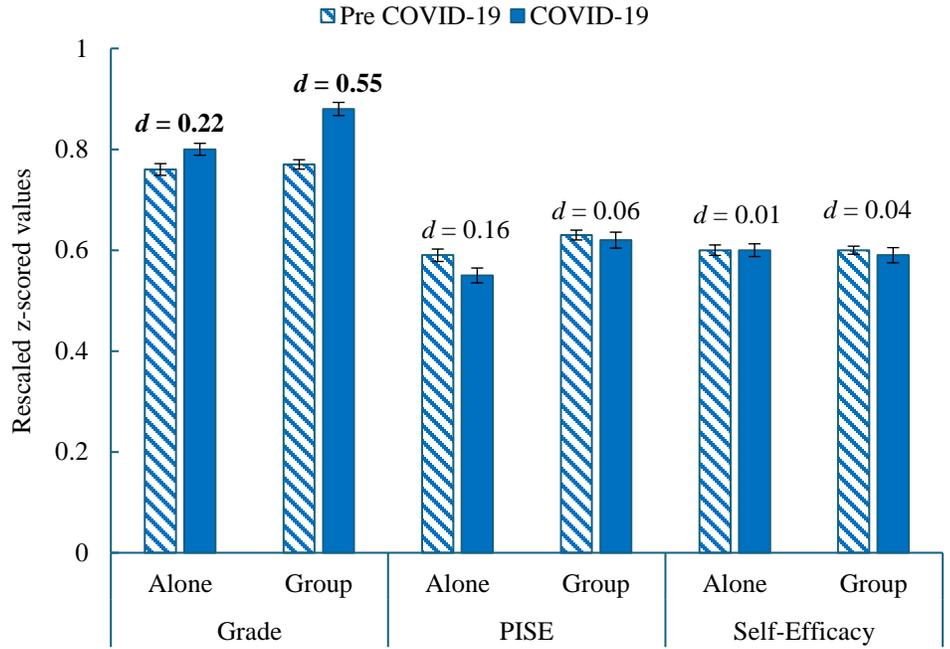

Figure A1. Comparison of the students who TWA (alone) before and during the pandemic and the students who TWG (group) before and during the pandemic (statistically significant values are bold).

**Appendix B**

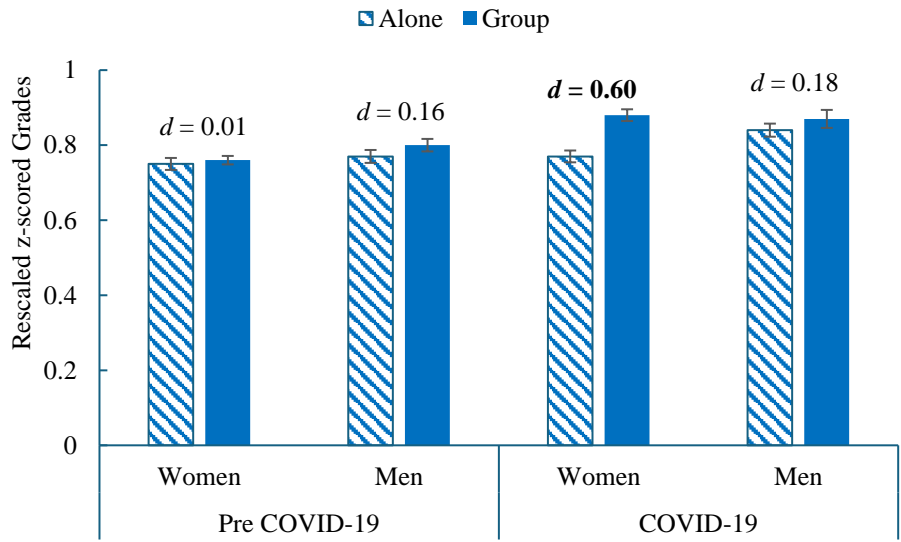

Figure A2. Grades of women and men who TWA (alone) / TWG (group) (statistically significant values are bold).

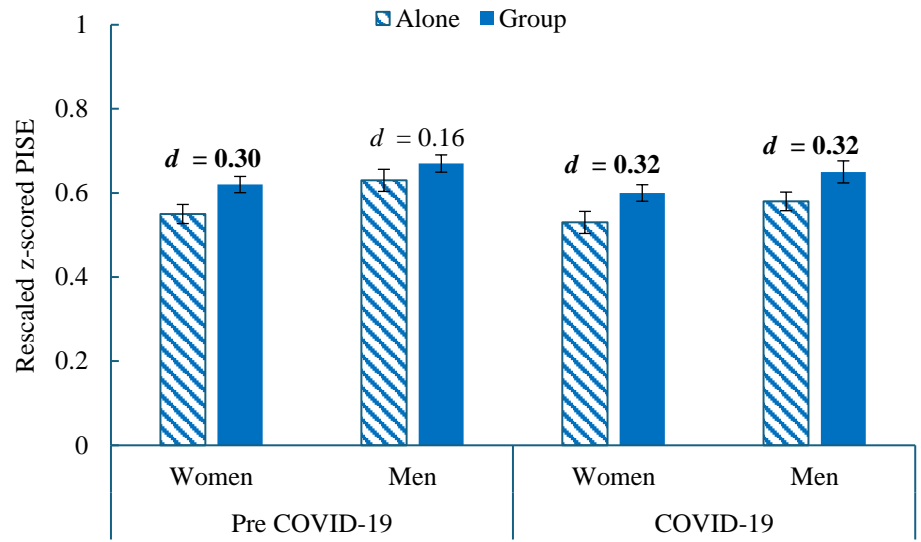

Figure A3. PISE of women and men who TWA (alone) / TWG (group) (statistically significant values are bold).

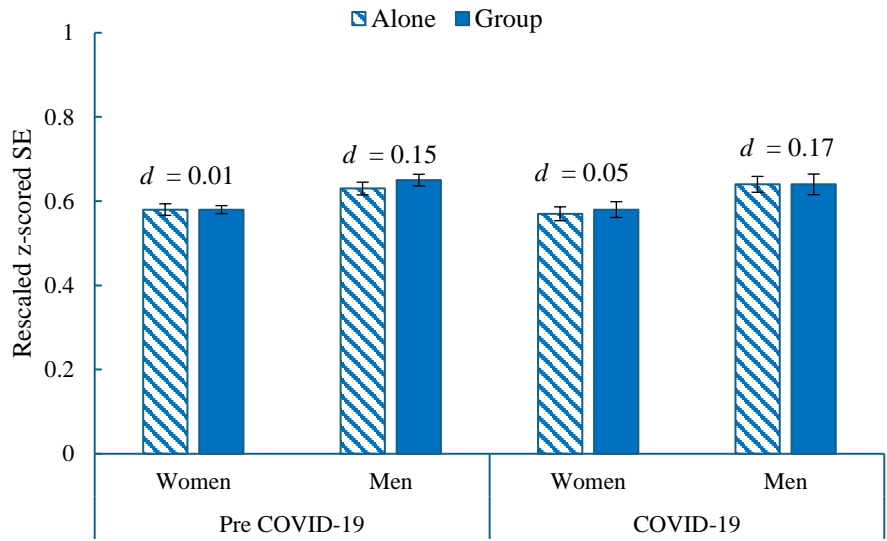

Figure A4. Physics self-efficacy of women and men who TWA (alone) / TWG (group) (statistically significant values are bold).